\documentstyle [12pt,epsfig]{article}
\title{Development of crystal extraction studies at the IHEP accelerator
    }

\author{V.M.Biryukov, V.N.Chepegin, Yu.A.Chesnokov,\\
V.I.Kotov, E.A.Lyudmirskiy, V.A.Maisheev,\\
E.F.Troyanov, N.K.Vishnevskiy, V.G.Zarucheiskiy\\
{\em\small Institute of High Energy Physics (Protvino), Russia} }

\date{Presented at PAC'97 (Vancouver, May 1997)}

\begin{document}
\maketitle

\begin{abstract}
The crystal extraction of 10$^7$ proton/s on BEC experiment facility
is described. Detailed computer analysis is made to explain the basic 
experimental results. Future high-intensity efficient extraction using a 
short crystal is announced.
\end{abstract}

In 1989 at the IHEP accelerator, a crystal
extraction of a 70-GeV proton beam onto the experimental set-up
PROZA[1] was carried out. As a development of this method,
another extraction  has been realized onto the experimental set-up BEC[2].
The beam line where BEC is located, was created for
formation from an internal target of the beams of negatively cherged
particles in the energy range of  20-40 GeV.

Usage in this case of the other methods of extraction of a proton beam
is extremely difficult, and requires a
significant reconstruction of an initial part of the beam line.
Using a bent
crystal, its bending angle and position inside
a vacuum chamber of the accelerator can be chosen so that the line of the
extraction of a proton beam is on the axis of the beam line.
An opportunity of extraction into the beam line of negatively
charged particles from internal targets is thus saved. The computations of
proton trajectories through a nonlinear magnetic
field of an accelerator were conducted under the programs FINT and TRAEK[3].
To not restrict the accelerator acceptance, the radial position of a
crystal was chosen equal to $\approx 50$ mm from the axis of the vacuum
chamber.
To bring protons onto the crystal,
a local distortion of a closed orbit was invoked.

 For extraction, a Si crystal of orientation
(111) with the sizes $ 85 \times 16 \times 0.5$ mm$^3 $, bent on a angle
of 89 mrad was used.
The goniometer, on which crystal was established, provided its
radial translation (coordinate accuracy of 0.1 mm) and turning
in a horizontal plane
with the step $\approx 80\mu$rad.

Up to 10\% of the intensity of an accelerated
  beam,  i.e. up to $10^{11}$ protons in a cycle, were incident
  on the crystal. Thus the intensity in the extraction beam line
was $10^7$ protons in a cycle  and, hence, the efficiency of extraction
was at the level $\sim 10^{-4}$ (see fig. 1).
The received intensity of protons was quite sufficient for fulfilment
of the planned experiment on BEC.
The low efficiency of extraction with the help of a crystal, bent on large
angle, as simulations show, is connected not only to intensive
dechanneling process of particles in such a crystal, but also to essential
influence of defects, inevitablly introduced at its manufacturing
(destroyed near-surface layer) and bending (twists).
The experiment was simulated by the program CATCH [4], which took into
account a geometry of the crystal with distortions and the effects of repeated
passage of particles through a crystal.
During the passage of particles through a bent crystal lattice, every step
($\sim$1 $\mu$m) the local crystal fields and densities of nuclei and
electrons were calculated, and the scattering events generated [4].
The lattice of a crystal was considered ideal, however on its  surface
there could be a nonchanneling layer of the thickness of a few microns
[5] ("septum thickness"), the influence of which was
investigated in details.
In the computation, the geometry of the holder of a crystal (scattering
in it) and variable longitudinal and transverse curvatures
of a crystal were taken into account.
Besides, the crystal twist was taken into account,
as a result of which the
orientation of atomic planes (111) at the entrance of a crystal becomes a
parabolic function of vertical coordinate $y$.
In our case an angle $\theta$ of misorientation of the planes (111)
is given by the expression $\theta (\mu$rad)=20$\times y^2$ (mm).
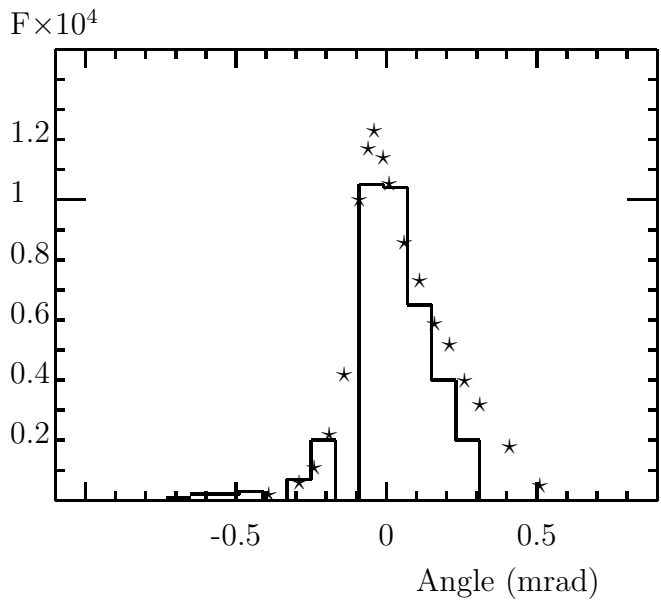
\begin{figure}[htb]
\begin{center}
\setlength{\unitlength}{2mm}
\begin{picture}(40,43)(-6,-7)
\thicklines
\linethickness{0.3mm}

\put(-5,0) {\line(1,0){40}}
\put(-5,0) {\line(0,1){30}}
\put(35,30) {\line(0,-1){30}}
\put(35,30) {\line(-1,0){40}}
\multiput(-5,0)(0,2){15}{\line(1,0){.6}}
\multiput(-5,20)(0,12){1}{\line(1,0){2}}
\multiput(35,20)(0,12){1}{\line(-1,0){2}}
\multiput(35,0)(0,2){15}{\line(-1,0){.6}}
\put(-8,24){\makebox(2,1)[l]{1.2}}
\put(-8,20){\makebox(2,1)[l]{1}}
\put(-8,16){\makebox(2,1)[l]{0.8}}
\put(-8,12){\makebox(2,1)[l]{0.6}}
\put(-8,8){\makebox(2,1)[l]{0.4}}
\put(-8,4){\makebox(2,1)[l]{0.2}}
\multiput(-5,0)(2,0){20}{\line(0,1){.6}}
\multiput(-3,0)(10,0){4}{\line(0,1){1.2}}
\multiput(-3,30)(10,0){4}{\line(0,-1){1.2}}
\multiput(-5,30)(2,0){20}{\line(0,-1){.6}}
\put(6,-3){\makebox(2,1)[b]{-0.5}}
\put(16,-3){\makebox(2,1)[b]{0}}
\put(26,-3){\makebox(2,1)[b]{0.5}}

\put(-8,31){F$\times$10$^4$}
\put(19,-6){Angle (mrad)}

\put(4.,.2) {\line(-1,0){1.6}}
\put(7.2,.4) {\line(-1,0){3.2}}
\put(7.2,.6) {\line(0,-1){.2}}
\put(8.8,.6) {\line(-1,0){1.6}}
\put(8.8,.6) {\line(0,-1){.6}}
\put(10.4,1.4) {\line(0,-1){1.4}}
\put(10.4,1.4) {\line(1,0){1.6}}
\put(12.,1.4) {\line(0,1){2.6}}
\put(12.,4) {\line(1,0){1.6}}
\put(13.6,4) {\line(0,-1){4}}
\put(15.2,21) {\line(0,-1){21}}
\put(15.2,21) {\line(1,0){1.6}}
\put(16.8,21) {\line(0,-1){0.2}}
\put(16.8,20.8) {\line(1,0){1.6}}
\put(18.4,20.8) {\line(0,-1){7.8}}
\put(18.4,13) {\line(1,0){1.6}}
\put(20.,8) {\line(0,1){5}}
\put(20.,8) {\line(1,0){1.6}}
\put(21.6,4) {\line(0,1){4}}
\put(21.6,4) {\line(1,0){1.6}}
\put(23.2,4) {\line(0,-1){4}}

\put(8,-0.1){ $\star$}
\put(10,.7){ $\star$}
\put(11,1.7){ $\star$}
\put(12,3.9){ $\star$}
\put(13,7.9){ $\star$}
\put(14,19.5){ $\star$}
\put(14.6,22.9){ $\star$}
\put(15,24.1){ $\star$}
\put(15.6,22.3){ $\star$}
\put(16,20.6){ $\star$}
\put(17,16.7){ $\star$}
\put(18,14.1){ $\star$}
\put(19,11.3){ $\star$}
\put(20,9.9){ $\star$}
\put(21,7.5){ $\star$}
\put(22,5.9){ $\star$}
\put(24,3.1){ $\star$}
\put(26,0.5){ $\star$}

\end{picture}
 \caption
  {Histogram is the measured dependence of the extraction efficiency
  on the crystal orientation angle. Points ($\star$) are the
  simulation results.
}               \label{u}
\end{center}
\end{figure}

The results of modeling are shown on figs. 1--3.
Computed for real conditions of the experiment, the efficiency of
extraction dependence on the angle of orientation of a crystal, shown on
fig. 1, well agrees with the results of the experiment. Modeling has shown,
that the chosen length of a crystal is optimum, while a vertical
gap of the
holder of 10 mm and accordingly the height of a crystal are not sufficient.
Indeed, in experiment the vertical size of undisturbed beam
was $\sim$13 mm.
Moreover, the inaccuracy of the middle plane  at the crystal position
is $\pm$5 mm.
Therefore the significant part of particles may be lost at the scattering
 on the holder.
If to increase a vertical gap of the holder of a crystal up to 20 mm,
the efficiency of extraction increases by 1.5-2 times.

Fig. 2 shows the computed
efficiency of extraction by an ideal crystal (no twist, nor near-surface
nonchanneling layer) as a function of the number $N$ of
particle's encounters with the crystal. The main contribution to efficiency
of extraction is from the first passage of particles through the crystal,
whereas the
 contribution of the subsequent passages quickly drops because of a large
 scattering of nonchanneled particles in a crystal.
The average number of passages of a particle through a crystal before
a capture into the channeling state, as follows from fig. 2,
is only $<N>$ =1.7.
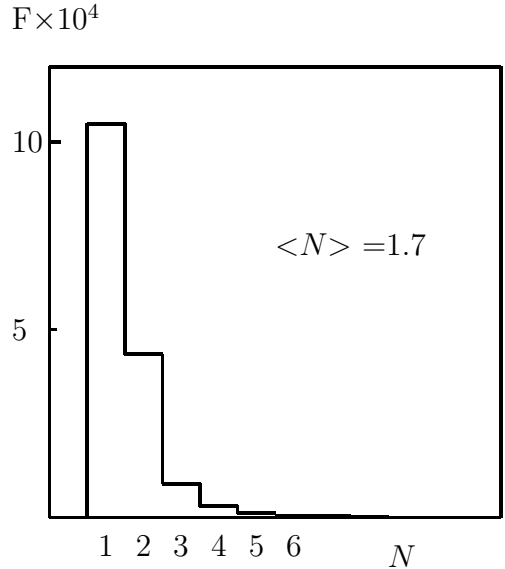
\begin{figure}[htb]
\begin{center}
\setlength{\unitlength}{5mm}
\begin{picture}(14,15)(-1,-1)
\thicklines
\linethickness{0.3mm}

\put(0,0) {\line(1,0){12}}
\put(0,0) {\line(0,1){12}}
\put(12,12) {\line(0,-1){12}}
\put(12,12) {\line(-1,0){12}}
\multiput(0,5)(0,5){1}{\line(1,0){.2}}
\multiput(0,10)(0,12){1}{\line(1,0){.3}}
\put(-1,9.5){\makebox(2,1)[l]{10}}
\put(-1,4.5){\makebox(2,1)[l]{5}}
\put(.5,-1){\makebox(2,1)[b]{1}}
\put(1.5,-1){\makebox(2,1)[b]{2}}
\put(2.5,-1){\makebox(2,1)[b]{3}}
\put(3.5,-1){\makebox(2,1)[b]{4}}
\put(4.5,-1){\makebox(2,1)[b]{5}}
\put(5.5,-1){\makebox(2,1)[b]{6}}

\put(-1,13){F$\times$10$^4$}
\put(9,-1.3){$N$}
\put(6,7){$<$$N$$>$ =1.7}

\put(1,0) {\line(0,1){10.48}}
\put(1,10.48) {\line(1,0){1}}
\put(2,10.48) {\line(0,-1){6.13}}
\put(2,4.35) {\line(1,0){1}}
\put(3,4.35) {\line(0,-1){3.45}}
\put(3,.90) {\line(1,0){1}}
\put(4,.90) {\line(0,-1){.59}}
\put(4,.31) {\line(1,0){1}}
\put(5,.31) {\line(0,-1){.19}}
\put(5,.12) {\line(1,0){1}}
\put(6,.06) {\line(1,0){1}}
\put(7,.04) {\line(1,0){1}}
\put(8,.02) {\line(1,0){1}}
\put(9,.01) {\line(1,0){1}}

\end{picture}
 \caption
  {The number of particles extracted by an ideal crystal at the
  $N$-th passage through the crystal.
}               \label{p}
\end{center}
\end{figure}

The presence of a near-surface nonchanneling layer, which
in our case has a thickness of the order 60 $\mu$m, and of a twist,
 changes this picture essentially.
If a beam is slowly brought onto a crystal, the primary impact parameter
is only a fraction of micron (the speed of this process is $\sim$5 mm/s,
the revolution time of a particle in a ring is 5 $\mu$s).
At such a depth, a particle hits a nonchanneling layer of a
crystal and, passing in it the way $\sim$1 á¬, scatters by $\sim$50
$\mu$rad, that results in a secondary impact parameter at the crystal
of $\sim$60 $\mu$m.
Thus, the presence in the crystal of the nonchanneling layer $\sim$60 $\mu$m
results in a complete suppression of efficiency of the first passage and
partial one of the second passage.
Dependence of efficiency of the extraction on a thickness of this
layer,  with twist and without it,
is shown in fig. 3.
\begin{figure}[htb]
\begin{center}
\setlength{\unitlength}{.5mm}
\begin{picture}(140,160)(-6,-12)
\thicklines
\linethickness{0.3mm}

\put(-5,0) {\line(1,0){110}}
\put(-5,0) {\line(0,1){180}}
\put(105,180) {\line(0,-1){180}}
\put(105,180) {\line(-1,0){110}}
\multiput(-5,0)(0,10){18}{\line(1,0){2}}
\multiput(-5,100)(0,1){1}{\line(1,0){4}}
\multiput(105,100)(0,1){1}{\line(-1,0){4}}
\multiput(105,0)(0,10){18}{\line(-1,0){2}}
\put(-15,160){\makebox(2,1)[l]{16}}
\put(-15,140){\makebox(2,1)[l]{14}}
\put(-15,120){\makebox(2,1)[l]{12}}
\put(-15,100){\makebox(2,1)[l]{10}}
\put(-15,80){\makebox(2,1)[l]{8}}
\put(-15,60){\makebox(2,1)[l]{6}}
\put(-15,40){\makebox(2,1)[l]{4}}
\put(-15,20){\makebox(2,1)[l]{2}}
\multiput(1,180)(2,0){51}{\line(0,-1){2}}
\multiput(1,180)(10,0){11}{\line(0,-1){4}}
\multiput(1.3,0)(2,0){51}{\line(0,1){2}}
\multiput(1.3,0)(10,0){11}{\line(0,1){4}}
\put(0,-7){\makebox(2,1)[b]{0}}
\put(20,-7){\makebox(2,1)[b]{20}}
\put(40,-7){\makebox(2,1)[b]{40}}
\put(60,-7){\makebox(2,1)[b]{60}}
\put(80,-7){\makebox(2,1)[b]{80}}
\put(100,-7){\makebox(2,1)[b]{100}}

\put(-12,184){F$\times$10$^4$}
\put(67,-12){Layer ($\mu$m)}

\put(0,45.){ \circle*{2}}
\put(1,28.7){ \circle*{2}}
\put(2,25.5){ \circle*{2}}
\put(5,24.6){ \circle*{2}}
\put(7,24.4){ \circle*{2}}
\put(10,22.5){ \circle*{2}}
\put(15,20.){ \circle*{2}}
\put(20,18.3){ \circle*{2}}
\put(30,15.4){ \circle*{2}}
\put(40,12.9){ \circle*{2}}
\put(50,11.2){ \circle*{2}}
\put(60,10.){ \circle*{2}}
\put(80,8.6){ \circle*{2}}
\put(100,8.0){ \circle*{2}}

\put(0,160.){ \circle{2}}
\put(1,80.){ \circle{2}}
\put(2,64.4){ \circle{2}}
\put(5,60.){ \circle{2}}
\put(7,56.5){ \circle{2}}
\put(10,52.2){ \circle{2}}
\put(15,43.6){ \circle{2}}
\put(20,36.1){ \circle{2}}
\put(30,28.4){ \circle{2}}
\put(40,22.){ \circle{2}}
\put(50,21.){ \circle{2}}
\put(60,19.6){ \circle{2}}
\put(80,18.3){ \circle{2}}
\put(100,17.4){ \circle{2}}

\end{picture}
 \caption
  {
 As simulated, the extraction efficiency as a function of the thickness
 of nonchanneling layer: ªcrystal with twist ($\bullet$) and without it (o).
}               \label{a}
\end{center}
\end{figure}
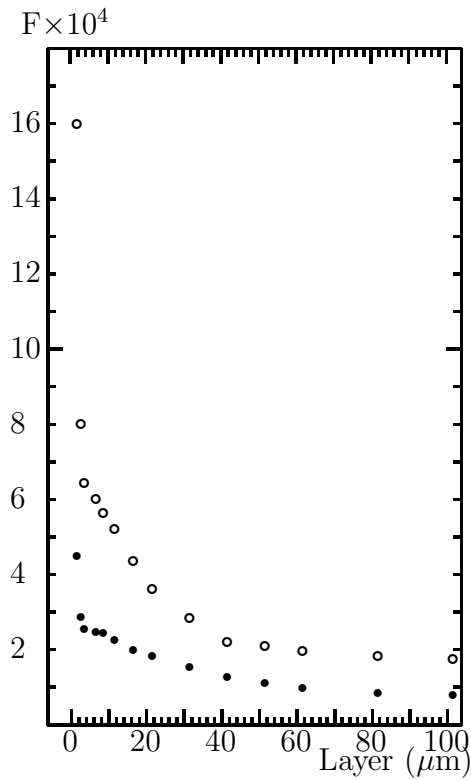

From there we see that twist
has a large influence on efficiency either.
The crystal with an ideal surface and without twist would ensure
an efficiency of the extraction one order of magnitude higher than presently.

A radical increase of efficiency of the extraction can be reached by a use of
a short crystal, bent on the small angle $\sim$ 1--3 mrad.
In a long, strongly-bent crystal the dechanneling losses are almost two
orders of magnitude.
Besides elimination of dechanneling losses of particles, the gain in
efficiency is reached also because of significant reduction of scattering
over the crystal length, i.e. respective reduction of the beam divergence
at the incidence on a crystal. Thus unlike in a long crystal, another
mechanism of the growth of efficiency of the extraction of particles
begins to work, related to the increase of the average number
of encounters of particles with a crystal.
Computations show that at the IHEP accelerator the efficiency of
extraction of
$\sim$ 20--40 \% by means of a crystal of the length of $\sim$5 mm,
bent on the angle $\sim$1.5 mrad, can be achieved.
The experiment on realization of such an extraction is in a stage of
preparation.

\end{document}